\documentstyle[prb,aps,multicol,eqsecnum]{revtex}

\renewcommand{\narrowtext}{\begin{multicols}{2}}
\renewcommand{\widetext}{\end{multicols}}

\begin{document}
\title{Ballistic electron motion in a random magnetic field.}
\author{K.B. Efetov$^{1,2}$ and V.R. Kogan${}^{1,2}$}
\address{${}^1$ Theoretische Physik III, \\
Ruhr-Universit\"at Bochum, 44780 Bochum, Germany\\
${}^2$ L. D. Landau Institute for Theoretical Physics, 117940 Moscow, Russia}
\date{\today}
\maketitle
\draft

\begin{abstract}
Using a new scheme of the derivation of the non-linear $\sigma $- model we
consider the electron motion in a random magnetic field (RMF) in two
dimensions. The derivation is based on writing quasiclassical equations and
representing their solutions in terms of a functional integral over
supermatrices $Q$ with the constraint $Q^{2}=1$. Contrary to the standard
scheme, neither singling out slow modes nor saddle-point approximations are
used. The $\sigma $-model obtained is applicable at the length scale down to
the electron wavelength. We show that this model differs from the model with
a random potential (RP). However, after averaging over fluctuations in the
Lyapunov region the standard $\sigma $-model is obtained leading to the
conventional localization behavior.
\end{abstract}

\pacs{PACS: 72.15.Rn, 73.20.Fz, 73.23.Ad}

\bigskip \narrowtext

\section{Introduction}

\label{introduction}

Description of the two dimensional ($2D$) electron motion in a random
magnetic field (RMF) is of a considerable interest for both experimentalists
and theoreticians. Two dimensional electron systems in a random magnetic
field were realized in a number of recent experiments when a high-mobility
heterostructure was located under an overlayer with randomly pinned flux
vortices in a type-II superconducting gate\cite{GBG92} or type-I
superconducting grains \cite{S94} or a demagnetized ferromagnet \cite{M95}.
From the theoretical point of view the RMF model is an example of a system
with the interaction which is realized through an effective gauge field. In
particular, this model arises in the theory of quantum Hall effect with a
half-filled Landau level \cite{HLR93}. Another application of this model is
a gauge field description of the doped Mott insulators \cite{IL89}.

One of the most important problems in the RMF models is the question about
localization of electron states. This question has been studied in many
numerical works and very different conclusions were drawn: from a) all the
states are localized, Refs.\cite{SN93}$^-$\cite{BSK98} to b) there may be a
band of delocalized states Refs.\cite{KWAZ93}$^-$\cite{SW00} and c) all the
states are localized except those with the precisely zero energy, Refs\cite
{MW96}$^{,}$ \cite{F99}. The problem of comparison of the results obtained
in different numerical calculations is a quite complicated task partly
because extended states and the states with very large localization length
can very often be hardly distinguished from each other.

From the point of view of the generally accepted scaling theory of
localization \cite{aalr} the RMF model should not be different from the
model describing the electron motion in a random potential in a homogeneous
magnetic field. In both the cases all electron states are expected to be
localized in $2D$ in an arbitrarily weak random potential. Using the
supersymmetry technique \cite{E83} this prediction was checked in several
works by deriving a proper $\sigma $-model. The authors of Ref. \cite{AMW94}
used the standard scheme of the derivation finding first the saddle-point in
the integral over supermatrices $Q$ and expanding then in slow modes near
this saddle point. As a result, they obtained a standard diffusive unitary $%
\sigma $-model similar to what one has for the model with a random potential
(RP) and the broken time reversal symmetry. The long range character of
correlations of the random vector potential, which is possible even if the
correlations of the magnetic field are short ranged, did not play any role.

A possibility of a new term in the $\sigma $-model due to special character
of the correlations of the vector potential was discussed later in Refs. 
\cite{te1,gm1,te2}. This was done by considering more carefully short
distances. A ballistic $\sigma $-model similar to that of Ref.\cite{MK95}
was derived in Refs.\cite{te1,te2} and the calculations were checked by
direct diagrammatic and path integrals methods \cite{gm1}. The final
conclusion of these works was that the $\sigma $-model maintained the
standard form \cite{AMW94} corresponding to the unitary ensemble unless the
correlations of the magnetic field were long ranged. This was considered, as
usual, as the proof of the localization. An additional term in the $\sigma $%
-model was still possible if the correlation of the magnetic field was
proportional to $q^{-2}$ Ref.\cite{te2}, where $q$ is the momentum, and this
could lead to antilocalization (see also Ref.\cite{kly}). However, no
possibility to obtain anything but the standard unitary $\sigma $-model and,
hence, the localization for any finite range correlations of the magnetic
field was seen finally from these works and no difference between the RMF
model and the RP model with a magnetic field was found even in the ballistic
case.

Nevertheless, the question about the localization in the RMF model \ in $2D$
was raised again in a recent numerical work \cite{N02}. On the basis of the
numerical study the author of Ref. \cite{N02} suggested quite a different
scenario of the electron motion in the RMF model \ arguing that there could
be some ``hidden degrees of freedom'' that lead to essential deviations from
the standard scaling description of disordered systems.

This result challenges the analytical results obtained on the basis of the $%
\sigma $-model description but it is fair to say that the previous
analytical study was not complete. All calculations were carried out using
the traditional form of the ballistic $\sigma $-model \cite{MK95,aos,te2}
with a conventional collision term. However, this form may be used for a
long range disorder at sufficiently long distances only. The derivation of
such a $\sigma $-model is based on finding a saddle point in the integral
over the supermatrices $Q$ and expanding in slow modes. This procedure fails
at short (but still much exceeding the wave length $\lambda _{F}$)
distances. As a result, the form of ballistic $\sigma $-model is not
applicable at the lengths smaller than a characteristic length $l_{L}\gg
\lambda _{F}$ and this puts doubts on some conclusions drawn previously. 

The saddle-point approximation is equivalent to the self-consistent Born
approximation (SCBA) and cannot be good for a long range disorder. At the
same time, even short range correlations of the magnetic field correspond to
long range correlations of the vector potential and this problem is
inevitably encountered in the RMF model. The diagrammatic expansion of Ref. 
\cite{gm1} also starts with the SCBA for one-particle Green functions and
one encounters the same problem.

In order to circumvent the problem related to the use of the saddle point
approximation and the expansion in the slow modes we suggested recently
another scheme \cite{EK03}. This method is based on equations for
quasiclassical Green functions and resembles the phenomenological approach
of Ref. \cite{MK95}. However, in contrast to the latter, we do not average
over disorder in the beginning of the calculations and do not decouple an
effective interaction by integration over an auxiliary field. Our approach
is exact in the quasiclassical limit and a resulting ballistic $\sigma $%
-model is applicable at all distances exceeding the wave length $\lambda
_{F} $. It can be reduced to the conventional ballistic $\sigma $-model
after a coarse graining procedure and the latter is applicable at distances
exceeding a Lyapunov length $l_{L}$ introduced in Ref. \cite{AL96}. At
distances smaller than $l_{L}$ the form of the term due to disorder is
different from the standard collision term.

In Ref. \cite{EK03} we derived the ballistic $\sigma $-model for the RP
models and now we present an analogous derivation for the RMF models. It
turns out that the terms in the ballistic $\sigma $-models describing the
disorder in the RP and RMF models differ from each other. They can become
similar only after carrying out the coarse graining procedure. We show that
this procedure can be performed in the same way as for the RP problem, which
leads to a similar reduced $\sigma $-model.

The paper is organized as follows: In the Chapter \ref{derivation} we
introduce a partition function generating correlation functions of interest
in terms of a functional integral over supervectors $\psi $. We derive
equations for Green function and simplify them using a quasiclassical
approximation. Introducing quasiclassical Green functions we rewrite the
equations in a gauge invariant form. The solution of the equations is found
in terms of an integral over supermatrices $Q$ with the constraint $Q^{2}=1$%
, which allows us to average over the RMF.

In the Chapter \ref{reduction} we integrate over fluctuations in the
Lyapunov region and come to a reduced $\sigma $- model with a collision term.

In the Appendix we consider the problem of the correlation of two particles
moving in a RMF and find the characteristic time of this correlation.

\section{Formulation of the problem. Quasiclassical approximation and
derivation of the $\protect\sigma $-model.}

\label{derivation} In the present work we follow the method of derivation of
the $\sigma $-model suggested in our previous work \cite{EK03}. In order to
make the presentation self-contained we repeat the main steps of the
derivation.

We start our consideration with the introduction of the partition function $%
Z[\hat{a}]$ 
\begin{equation}
Z[\hat{a}]=\int \exp \left( -L_{a}[\psi ]\right) D\psi  \label{eq1}
\end{equation}
\begin{eqnarray}
L_{a}[\psi ] &=&-i\int \bar{\psi}({\bf r})\left( \hat{H}({\bf r}%
)-\varepsilon +\frac{\omega }{2}+\frac{\omega +i\delta }{2}\Lambda \right)
\psi ({\bf r})d{\bf r}+  \nonumber \\
&&+i\int \bar{\psi}({\bf r})\hat{a}({\bf r})\psi ({\bf r})d{\bf r}  \nonumber
\end{eqnarray}
where $\psi $ are $8$-component supervectors \cite{E83} and the Hamiltonian $%
\hat{H}({\bf r})$ in Eq.(\ref{eq1}) is taken in the form 
\begin{equation}
\hat{H}({\bf r})=\left( -i{\bf \nabla _{r}}-\frac{e}{c}\hat{\tau}_{3}{\bf A}(%
{\bf r})\right) ^{2}/2m-\varepsilon _{F}+u({\bf r})  \label{eq2}
\end{equation}
The last term in Eq.(\ref{eq1}) contains a source function $\hat{a}({\bf r})$%
. Choosing this function in a proper form and taking derivative in it one
can obtain correlation functions. For example, the level-level correlation
function $R(\omega )$ can be written as: 
\begin{equation}
R(\omega )=\frac{1}{2}-\frac{1}{2(\pi \nu V)^{2}}\lim_{\alpha _{1}=\alpha
_{2}=0}{\rm Re}\frac{\partial ^{2}}{\partial \alpha _{1}\partial \alpha _{2}}%
Z[\hat{a}]  \label{eq3}
\end{equation}
where the source $\hat{a}({\bf r})$ is the following matrix: 
\begin{equation}
\hat{a}({\bf r})=\left( 
\begin{array}{ccc}
\hat{\alpha}_{1} & 0 &  \\ 
0 & -\hat{\alpha}_{2} & 
\end{array}
\right) ,\text{ \ \ }\hat{\alpha}_{1,2}=\frac{\alpha _{1,2}}{2}(1-k)
\label{eq4}
\end{equation}
Here $k$ is the diagonal matrix with elements $\pm 1$ in fermionic and
bosonic blocks respectively \cite{E83}.

The Hamiltonian $H({\bf r})$, Eq.(\ref{eq2}), contains both scalar and
vector potentials $u({\bf r})$, ${\bf A}({\bf r})$ that are assumed to be
random functions of the space coordinates distributed according to the Gauss
law, $\hat{\tau}_{3}$ is the third Pauli matrix in the particle-hole space.
Below we consider a general case when the scalar potential $u({\bf r})$
contains both the short range $u_{s}\left( {\bf r}\right) $ and long range $%
u_{l}({\bf r})$ parts with the characteristic correlation lengths of the
order and larger than the Fermi wavelength $\lambda _{F}=(2\pi p_{F})^{-1}$
respectively. Their statistics are determined by the pair correlation
functions: 
\begin{eqnarray}
&&\langle u_{s}({\bf r})u_{s}({\bf r^{\prime }})\rangle =\frac{1}{2\pi \nu
\tau _{s}}\delta ({\bf r}-{\bf r^{\prime }})  \label{eq5} \\
&&\langle u_{l}({\bf r})u_{l}({\bf r^{\prime }})\rangle =W({\bf r}-{\bf %
r^{\prime }})  \label{eq6}
\end{eqnarray}
where the function $W({\bf r}-{\bf r^{\prime }})$ is assumed to fall off
over a length $d\gg \lambda _{F}$. Statistics of the magnetic field will be
introduced later. Although the main goal of this paper is to study the RMF
model, we add the scalar potential into the Hamiltonian for a more explicit
comparison between the RMF and RP models.

Following the standard approach of Ref.\cite{E83} one would average the
partition function $Z[\hat{a}]$, Eq.(\ref{eq1}), over the random external
fields and then, singling out fluctuations slowly varying in space and
integrating over an auxiliary smooth matrix field $Q$, decouple the
interaction term $(\psi \bar{\psi})^{2}$ that appears after the averaging.
This method was recently used, e.g., in Ref.\cite{AAAS96} in a derivation of
the ballistic $\sigma $-model for quantum billiards and in Ref.\cite{te1,te2}%
, where the two-dimensional electron gas was considered in a random magnetic
field. As it has been mentioned in the section \ref{introduction} the latter
problem is rather specific because the vector potential ${\bf A}({\bf r})$
can have long range correlations even if correlations of the magnetic field
are short ranged.

The singling out of slow modes with the subsequent decoupling of the
interaction by integrating over an auxiliary smooth matrix $Q$ is not a
rigorous procedure because some part of the interaction is assumed to be
irrelevant and is neglected. Although this assumption works well for short
range impurities, it is not justified for long range correlations. Below we
use another method based on the Green function and quasiclassical
approximation of Ref. \cite{EK03}. This method allows one to derive a $%
\sigma $-model applicable down to the length scale of the order of the
wavelength $\lambda _{F}$.  

Following Ref. \cite{EK03} we average over the short range potential $u_{s}(%
{\bf r})$, decouple the interaction term appearing after this averaging
using the standard integration over an auxiliary smooth matrix field $M({\bf %
r})$ and finally rewrite the partition function as follows: 
\begin{equation}
Z[\hat{a}]=\int Z_{1}[J]\exp \left( -\frac{\pi \nu }{8\tau _{s}}\int {\rm Str%
}M^{2}({\bf r})d{\bf r}\right) DM  \label{eq7}
\end{equation}
where 
\begin{equation}
Z_{1}[J]=\int \exp (-L_{J}[\psi ])D\psi   \label{eq8}
\end{equation}
The Lagrangian $L_{J}[\psi ]$ coincides with $L_{a}[\psi ]$, Eq.(\ref{eq1}),
provided the substitutions $u_{s}({\bf r})=0$ and $i\hat{a}({\bf r}%
)\rightarrow J({\bf r})=i\hat{a}({\bf r})+M({\bf r})/2\tau _{s}$ are made in
the Lagrangian $L_{a}[\psi ]$, Eq.(\ref{eq1}). The structure of the matrix $%
M({\bf r})$ can be found in the book, Ref.\cite{E83}. It is important that $%
M({\bf r})$ is self-conjugate: $\bar{M}({\bf r})=M({\bf r})$ where the bar
means the ``charge conjugation'' 
\[
\bar{M}({\bf r})=CM^{T}({\bf r})C^{T}
\]
\[
C=\Lambda \otimes \left( 
\begin{array}{cc}
c_{1} & 0 \\ 
0 & c_{2}
\end{array}
\right) ,\text{ \ }c_{1}=\left( 
\begin{array}{cc}
0 & -1 \\ 
1 & 0
\end{array}
\right) ,\text{ \ }c_{2}=\left( 
\begin{array}{cc}
0 & 1 \\ 
1 & 0
\end{array}
\right) 
\]
(see also Ref.\cite{E83}). 

Following Refs.\cite{MK95,EK03} we introduce the Green function $G({\bf r},%
{\bf r^{\prime }})$ 
\begin{equation}
G^{\alpha \beta }({\bf r},{\bf r^{\prime }})=Z_{1}^{-1}[J]\int \psi ^{\alpha
}({\bf r})\bar{\psi}^{\beta }({\bf r^{\prime }})e^{-L_{J}[\psi ]}D\psi 
\label{eq9}
\end{equation}
For the most correlation functions of interest the source function $\hat{a}(%
{\bf r})$ can be chosen to be self-conjugate. If this is the case the Green
function satisfies the equation 
\begin{equation}
\left[ \hat{H}({\bf r})-\varepsilon +\frac{\omega }{2}+\frac{\omega +i\delta 
}{2}\Lambda +iJ({\bf r})\right] G({\bf r},{\bf r^{\prime }})=i\delta ({\bf r}%
-{\bf r^{\prime }})  \label{eq10}
\end{equation}
Eq.(\ref{eq10}) was previously studied in the absence of the magnetic field
in the quasiclassical approximation using a method of a quasiclassical Green
function, Refs.\cite{MK95,EK03}. This method is based on the assumption that
the external fields and sources are smooth functions (i.e. slowly changing
over the wavelength $\lambda _{F}$). Within this method the Green function $%
G({\bf p},{\bf R})$ can be rewritten using the Wigner transformation 
\[
G({\bf r},{\bf r^{\prime }})=\int \frac{d{\bf p}}{(2\pi )^{2}}e^{i{\bf p}(%
{\bf r}-{\bf r^{\prime })}}G({\bf p},{\bf R}),\text{ \ \ }{\bf R}=({\bf r}+%
{\bf r^{\prime }})/2
\]
The function $G\left( {\bf p,R}\right) $ has a sharp peak at the Fermi
surface ${\bf p}=p_{F}{\bf n}$. This property is due to the fact that the
long range fields and sources weakly disturb the shape of the Fermi surface.
Integrating the Green function $G({\bf p},{\bf R})$ over the absolute value
of the momentum ${\bf p}$ results in a new function $g_{{\bf n}}({\bf r})$
that depends on the centre of mass coordinate ${\bf R}$ and the unit vector $%
{\bf n}={\bf p}/p$ determining the direction at the Fermi surface. The
coordinate dependence of this function turns out to be smooth and therefore $%
g_{{\bf n}}({\bf r})$ may be considered as the quasiclassical approximation
of the exact Green function $G({\bf r},{\bf r^{\prime }})$. On the other
hand, the partition function $Z_{1}[J]$, Eq.(\ref{eq8}), can be expressed
through $g_{{\bf n}}({\bf r})$.

Before we start the calculation following this procedure let us make some
remarks about differences between the RP and RMF models. First, the presence
of the magnetic field breaks the time-reversal symmetry and, hence,
excitations sensitive to the time reversal are suppressed. Therefore we
consider only such correlation functions that can be obtained from the
sources $\hat{a}({\bf r})$ commuting with $\tau _{3}$. The part of the Green
function anticommuting with $\tau _{3}$ is negligible and may be omitted
from the further consideration.

The second remark is related to the physical aspects of the quasiclassical
approximation in the presence of a magnetic field. It is known that systems
placed in a magnetic field are invariant with respect to the magnetic
translations $\hat{T}_{{\bf a}}=\exp \left[ \left( {\bf \nabla _{r}}-i(e/c)%
\hat{\tau}_{3}{\bf A}\right) {\bf a}\right] $ instead of the ordinary ones 
\cite{LLIX}. The difference between these translations is relevant for an
infinite system even if the magnetic field is weak. This means that electron
states are to be characterized not by the ordinary momentum ${\bf p}_{kin}$
determining the kinetic energy but rather by the generalized momentum ${\bf p%
}={\bf p}_{kin}+(e/c)\hat{\tau}_{3}{\bf A}({\bf r})$. The generalized
momentum ${\bf p}$ is a well-defined quantum number if the magnetic field is
weak: 
\begin{equation}
r_{H}\gg \lambda _{F},\text{ \ \ }r_{H}=\frac{v_{F}}{\omega _{H}}
\label{eq11}
\end{equation}
where $\omega _{H}=eH/mc$ is the Larmor frequency and $v_{F}$- Fermi
velocity. Inequality (\ref{eq11}) coincides with the condition of the
applicability of the quasiclassical approximation. The Fermi surface is
defined in the space of the generalized momentum ${\bf p}$ and, contrary to
the case of zero magnetic field, has a rather complicated form. The value of
the momentum ${\bf p}$ at the Fermi surface strongly depends on the
direction ${\bf n}={\bf p}/p$. Therefore we change the definition of the
quasiclassical Green function by replacing the integration over the absolute
value of the generalized momentum ${\bf p}$ by that of the kinetic one ${\bf %
p}_{kin}$ (see e.g. Refs.\cite{K65}): 
\begin{equation}
g_{{\bf n}}({\bf r})=\frac{1}{\pi }\int d\xi G\left( {\bf p}+\frac{e}{c}\hat{%
\tau}_{3}{\bf A}({\bf r}),{\bf r}\right)   \label{eq12}
\end{equation}
where the function $G({\bf p},{\bf r})$ is the Green function taken in the
Wigner representation and $\xi ={\bf p}^{2}/(2m)-\varepsilon _{F}$, ${\bf n}=%
{\bf p}/p$. The quasiclassical Green function $g_{{\bf n}}({\bf r})$ defined
by Eq.(\ref{eq12}) is gauge-invariant. The logarithmic derivative of the
partition function $Z_{1}[J]$, Eq.(\ref{eq8}), can be estimated as follows: 
\begin{equation}
\frac{\delta \ln Z_{1}[J]}{\delta J({\bf r})}=\frac{1}{2}G({\bf r},{\bf r}%
)\approx \frac{\pi \nu }{2}\int g_{{\bf n}}({\bf r})d{\bf n}  \label{eq13}
\end{equation}
where $\nu $ is the density of states at the Fermi surface. Performing the
Wigner transformation we subtract Eq.(\ref{eq10}) from the conjugated one,
then integrate the result over $\xi $ as in the Eq.(\ref{eq12}) and obtain
in the quasiclassical approximation: 
\[
\left( v_{F}{\bf n\nabla _{r}}+\frac{e}{mc}\hat{\tau}_{3}B({\bf r})\partial
_{\varphi }-p_{F}^{-1}{\bf \nabla _{r}}u({\bf r}){\bf \partial _{n}}\right)
g_{{\bf n}}({\bf r})+
\]
\begin{equation}
\frac{i(\omega +i\delta )}{2}[\Lambda ,\;g_{{\bf n}}]-[J({\bf r}),\;g_{{\bf n%
}}]=0  \label{eq14}
\end{equation}
In Eq. (\ref{eq14}), $B({\bf r})=\partial _{x}A_{y}-\partial _{y}A_{x}$ is
the magnetic field, ${\bf \partial _{n}}={\bf e}_{\varphi }\partial
_{\varphi }-{\bf n}$, ${\bf e}_{\varphi }=(-\sin \varphi ,\;\cos \varphi )$.
In this approximation the solution of the Eq.(\ref{eq14}) is to be sought
with the usual constraint \cite{EK03} 
\begin{equation}
g_{{\bf n}}^{2}({\bf r})=1  \label{eq15}
\end{equation}
and the boundary condition 
\begin{equation}
\left. g_{{\bf n}_{\perp }}({\bf r})=g_{-{\bf n}_{\perp }}({\bf r})\right| _{%
{\bf r}\in S}  \label{eq16}
\end{equation}
where ${\bf r}\in S$ stands for points on the surface of the sample and $%
{\bf n}_{\perp }$ means the component of the vector ${\bf n}$ perpendicular
to the surface. Following Ref. \cite{EK03} we write the solution of Eq.(\ref
{eq14}) in terms of a functional integral over supermatrices $Q_{{\bf n}%
}\left( {\bf r}\right) $  
\[
g_{{\bf n}}({\bf r})=Z_{2}^{-1}[J]\int_{Q_{{\bf n}}^{2}=1}Q_{{\bf n}}({\bf r}%
)\exp \left( -\frac{\pi \nu }{2}\Phi _{J}[Q_{{\bf n}}]\right) DQ_{{\bf n}}
\]
\begin{equation}
Z_{2}[J]=\int_{Q_{{\bf n}}^{2}=1}\exp \left( -\frac{\pi \nu }{2}\Phi _{J}[Q_{%
{\bf n}}]\right) DQ_{{\bf n}}  \label{eq17a}
\end{equation}
\begin{eqnarray}
&&\Phi _{J}[Q_{{\bf n}}]={\rm Str}\int d{\bf r}d{\bf n}\biggl[\Lambda \bar{T}%
_{{\bf n}}({\bf r})\biggl(v_{F}{\bf n\nabla _{r}}+\frac{eB({\bf r})}{mc}\hat{%
\tau}_{3}\partial _{\varphi }-  \nonumber \\
&&p_{F}^{-1}{\bf \nabla _{r}}u({\bf r}){\bf \nabla _{n}}\biggr)T_{{\bf n}}(%
{\bf r})+\left( \frac{i(\omega +i\delta )}{2}\Lambda -J({\bf r})\right) Q_{%
{\bf n}}({\bf r})\biggr]  \label{eq17}
\end{eqnarray}
\[
Q_{{\bf n}}({\bf r})=T_{{\bf n}}({\bf r})\Lambda \bar{T}_{{\bf n}}({\bf r}),%
\text{ \ \ }\bar{T}_{{\bf n}}({\bf r})T_{{\bf n}}({\bf r})=1
\]
where $\partial _{\varphi }$ stands for the derivative in the angle. The
integration in Eq.(\ref{eq17}) is performed over the self-conjugate
supermatrices 
\[
\bar{Q}_{{\bf n}}({\bf r})=Q_{{\bf n}}({\bf r}),\text{ \ \ }\bar{Q}_{{\bf n}%
}({\bf r})=CQ_{-{\bf n}}^{T}({\bf r})C^{T}
\]
with the constraint $Q_{{\bf n}}^{2}({\bf r})=1$ and 
\begin{equation}
Q_{{\bf n}_{\perp }}({\bf r})|_{S}=Q_{-{\bf n}_{\perp }}({\bf r})|_{S}
\label{eq18}
\end{equation}
at the surface $S$ of the sample. The structure of the supermatrix $Q_{{\bf n%
}}$ coincides with the structure of the supermatrix $M({\bf r})$. We do not
demonstrate here the equivalence of the matrices $g_{{\bf n}}\left( {\bf r}%
\right) $, Eqs.(\ref{eq12}), (\ref{eq17}), and refer to the proof given in
Ref.\cite{EK03}. We mention here only that both the matrices are the
logarithmic derivatives in the matrix $J({\bf r})$ of the partition
functions $Z_{1}[J]$, $Z_{2}[J]$ respectively. Hence, these functions are
equal to each other up to some factor that is independent of $J({\bf r})$.
Due to the supersymmetry $Z_{1}[J]=Z_{2}[J]=1$ for $J({\bf r})=0$, which
means that the factor is unity and the partition functions are equal to each
other 
\begin{equation}
Z_{1}[J]=Z_{2}[J]  \label{eq19}
\end{equation}

Below the magnetic field $B({\bf r})$ is considered as a random function
with a Gaussian distribution and the pair correlation function of the form 
\begin{equation}
\langle B({\bf r})B({\bf r^{\prime }})\rangle =2\left( \frac{mc}{e}\right)
^{2}\omega _{c}^{2}W_{B}({\bf r}-{\bf r^{\prime }})  \label{eq20}
\end{equation}
where is $\ \omega _{c}$ is a coefficient that has a meaning of the
characteristic frequency of the cyclotron motion and the function $W_{B}(%
{\bf r}-{\bf r^{\prime }})$ is assumed to fall off at distances $|{\bf r}-%
{\bf r^{\prime }}|>b$ and to be normalized as $W_{B}({\bf r}=0)=1$. The
length $b$ characterizes the decay of the correlations of the RMF $B\left( 
{\bf r}\right) $. Substituting Eq.(\ref{eq19}) into Eq.(\ref{eq7}) and
averaging the result over the magnetic field and long-ranged potential $%
u_{l}({\bf r})$ we find for the partition function $Z[\hat{a}]$ Eq.(\ref{eq7}%
) 
\begin{equation}
Z[\hat{a}]=\int \exp (-F[Q_{{\bf n}}])DQ_{{\bf n}}  \label{eq21}
\end{equation}
where the free energy functional $F[Q_{{\bf n}}]$ has the form: 
\[
F[Q_{{\bf n}}]=F_{kin}[Q_{{\bf n}}]+F_{imp}[Q_{{\bf n}}]+F_{imp}^{(s)}[Q_{%
{\bf n}}]+F_{m}[Q_{{\bf n}}] 
\]
\begin{eqnarray}
F_{kin}[Q_{{\bf n}}] &=&\frac{\pi \nu }{2}{\rm Str}\int d{\bf r}d{\bf n}%
\biggl[\Lambda {\bar{T}}_{{\bf n}}({\bf r})v_F{\bf n\nabla_r}T_{{\bf n}}(%
{\bf r})  \nonumber \\
&&+i\left( \frac{\omega +i\delta }{2}\Lambda -\hat{a}\right) Q_{{\bf n}}(%
{\bf r})\biggr]  \nonumber
\end{eqnarray}
\begin{eqnarray}
&&F_{imp}\left[ Q_{{\bf n}}\right] =-\frac{1}{8}\left( \frac{\pi \nu }{p_{F}}%
\right) ^{2}\int d{\bf r}d{\bf n}d{\bf r}{^{\prime }}d{\bf n}{^{\prime }}%
{\bf \nabla }_{{\bf r}}^{i}{\bf \nabla }{}_{{\bf r}^{\prime }}^{j}W({\bf r}-%
{\bf r}{^{\prime }})  \nonumber \\
&&\times {\rm Str}[\Lambda {\bar{T}}_{{\bf n}}\left( {\bf r}\right) \nabla _{%
{\bf n}}^{i}T_{{\bf n}}({\bf r})]{\rm Str}[\Lambda {\bar{T}}_{{\bf n}{%
^{\prime }}}({\bf r}{^{\prime }})\nabla _{{\bf n}{^{\prime }}}^{j}T_{{\bf n}{%
^{\prime }}}({\bf r}{^{\prime }})]  \label{eq22}
\end{eqnarray}
\[
F_{imp}^{\left( s\right) }\left[ Q_{{\bf n}}\right] =-\frac{\pi \nu }{8\tau
_{s}}\int{\rm Str}\left( \int Q_{{\bf n}}\left( {\bf r}\right) d{\bf n}%
\right) ^{2}d{\bf r} 
\]
\begin{eqnarray}
&&F_{m}[Q_{{\bf n}}]=\left( \frac{\pi \nu }{2}\omega _{c}\right) ^{2}\int d%
{\bf r}d{\bf n}d{\bf r^{\prime }}d{\bf n^{\prime }}W_{B}({\bf r}-{\bf %
r^{\prime }})  \nonumber \\
&&\times {\rm Str}\left( \Lambda \hat{\tau}_{3}\bar{T}_{{\bf n}}({\bf r}%
)i\partial _{\varphi }T_{{\bf n}}({\bf r})\right) {\rm Str}\left( \Lambda 
\hat{\tau}_{3}\bar{T}_{{\bf n^{\prime }}}({\bf r^{\prime }})i\partial
_{\varphi ^{\prime }}T_{{\bf n^{\prime }}}({\bf r^{\prime }})\right) 
\nonumber
\end{eqnarray}
and 
\begin{equation}
\nabla _{{\bf n}}=-[{\bf n}\times \lbrack {\bf n}\times \frac{\partial }{%
\partial {\bf n}}]]={\bf e}_{\varphi}\partial_{\varphi}  \label{eq22a}
\end{equation}

The first term $F_{kin}[Q_{{\bf n}}]$ describes the free motion and is what
remains when external fields and impurities are absent. The second and the
third terms $F_{imp}\left[ Q_{{\bf n}}\right] $, $F_{imp}^{(s)}[Q_{{\bf n}}]$
are responsible for the scattering on the long- and short-ranged potentials
respectively. The last term $F_{m}\left[ Q_{{\bf n}}\right] $ is due to the
presence of the random magnetic field. Correlation functions of interest can
be obtained by calculating derivatives in the source $\hat{a}({\bf r})$ of
the partition function $Z[\hat{a}]$, Eq.(\ref{eq21}).

It is important to emphasize that the structure of the terms $F_{imp}[Q_{%
{\bf n}}]$ and $F_{m}[Q_{{\bf n}}]$ describing the electron scattering on
the random potential and on the random magnetic field, respectively, is
clearly different. The term $F_{imp}[Q_{{\bf n}}]$ contains the components
of the gradients parallel to the plane, whereas the term $F_{m}[Q_{{\bf n}}]$
contains the perpendicular one.

Nevertheless, at longer distances the RP and the RMF models are very similar
and we show this in the next Chapter carrying out a coarse graining
procedure suggested in Ref.\cite{EK03}. The latter means integrating out
degrees of freedom at distances inside the Lyapunov region.

For simplicity of the presentation we will consider in the next Chapters
only effects related to the random magnetic field and disregard the
scattering on the random potentials omitting $F_{imp}\left[ Q_{{\bf n}}%
\right] $, $F_{imp}^{(s)}[Q_{{\bf n}}]$ in the free energy Eq.(\ref{eq22}).
Accordingly, we will consider the symmetry of the supermatrices $Q$
corresponding to the unitary ensemble. We will study the behavior of the $%
\sigma $-model, Eq.(\ref{eq22}), on different length scales and discuss the
connection of this model with the models previously obtained in Refs.\cite
{AMW94,te1,gm1,te2}.

\section{Reduced $\protect\sigma $-model}

\label{reduction}

The $\sigma $-model obtained in Eq.(\ref{eq22}) is valid for the length
scales down to the wavelength $\lambda _{F}$ and has the form which differs
from the $\sigma $- model found in the Ref.\cite{te1,te2}. The latter model
has been derived for the spatially uncorrelated magnetic field and is
applicable at the length scale restricted from below by the single-particle
relaxation length $l$ but not by the wavelength $\lambda _{F}$. The length $%
l $ could not be consistently estimated within the consideration of Refs. 
\cite{te1,te2} and remained without a clear physical interpretation. At the
same time, the analysis of Refs.\cite{AL96,GM02,EK03} leads to the
conclusion that the role of this length is played by the Lyapunov length $%
l_{L}=v_{F}\tau _{L}$. Here $\tau _{L}$ is the inverse Lyapunov exponent and
is the time during which two close trajectories increase the distance
between them by a factor of the order of unity. On the other hand, according
to the Ref.\cite{AL96}, $\tau _{L}$ is the time which is required for two
scattered particles to diverge over the distance of the order of the range
of the potential (or the correlation length). In the Appendix we discuss the
problem of the particle motion in a RMF and estimate the Lyapunov length $%
l_{L}$ for weak fields as 
\begin{equation}
l_{L}\sim l_{tr}\left( \frac{b}{l_{tr}}\right) ^{2/3}  \label{e3.1}
\end{equation}
This result shows that the Lyapunov length $l_{L}$ is between the
correlation $b$ and transport $l_{tr}$ lengths: $b\ll l_{L}\ll l_{tr}$.

The Lyapunov length $l_{L}$ divides the length scales into two regions. At
small distances, two particles propagate in the same magnetic field and
correlations between them are relevant. Following the terminology of Ref. 
\cite{AL96} we call these distances the Lyapunov region. In the second
region when the scales of interest are larger than the Lyapunov length, the
motion of the particles is not correlated and they are scattered by the RMF
independently. This can be called the collision region because the
corresponding classical motion at such distances is described by the
conventional Boltzmann equation with a collision term corresponding to the
scattering on the RMF. The electron motion at these long distances should be
described by a reduced $\sigma $-model and one can expect that this reduced $%
\sigma $-model is just the $\sigma $-model of Ref.\cite{te1,te2}. In order
to obtain the reduced $\sigma $-model one should integrate out in Eqs. (\ref
{eq21}, \ref{eq22}) the degrees of freedom related to the Lyapunov region.
This coarse graining procedure has been worked out in Ref.\cite{EK03} for
the RP model and we will repeat it now for the RMF model.

First, one should explicitly decouple the original mode $T_{{\bf n}}({\bf r}%
) $ into the ''slow'' and ''fast'' parts. We make this separation in the way
preserving the rotational invariance of the initial model Eq.(\ref{eq22}): 
\begin{equation}
T_{{\bf n}}({\bf r})=\tilde{T}_{{\bf n}}({\bf r})V_{{\bf n}}({\bf r})
\label{eq23}
\end{equation}
Here $\tilde{T}_{{\bf n}}({\bf r})$, $V_{{\bf n}}({\bf r})$ are ''slow'' and
''fast'' modes describing the fluctuations in the collision and Lyapunov
regions respectively. As soon as the mode separation is made one should
substitute Eq.(\ref{eq23}) into the free energy $F[Q_{{\bf n}}]$, Eq.(\ref
{eq22}), and then average it over the ''fast'' fluctuations $V_{{\bf n}}(%
{\bf r})$: 
\begin{equation}
Z[\hat{a}]=\int_{\tilde{Q}_{{\bf n}}^{2}=1}e^{-F_{eff}[\tilde{Q}_{{\bf n}}]}D%
\tilde{Q}_{{\bf n}}  \label{eq24}
\end{equation}
where 
\begin{equation}
e^{-F_{eff}[\tilde{Q}_{{\bf n}}]}=\int \exp (-F[Q_{{\bf n}%
}^{(0)}]-F_{int}[Q_{{\bf n}}^{(0)},\tilde{Q}_{{\bf n}}])DV_{{\bf n}}
\label{eq25}
\end{equation}
\[
Q_{{\bf n}}^{(0)}({\bf r})=V_{{\bf n}}({\bf r})\Lambda \bar{V}_{{\bf n}}(%
{\bf r}),\text{ \ \ }\tilde{Q}_{{\bf n}}({\bf r})=\tilde{T}_{{\bf n}}({\bf r}%
)\Lambda \bar{\tilde{T}}_{{\bf n}}({\bf r}) 
\]
The functional $F[Q_{{\bf n}}^{(0)}]$ in Eq.(\ref{eq25}) coincides with the
free energy Eq.(\ref{eq22}) provided the source is omitted in the latter
expression. The functional $F_{int}[Q_{{\bf n}}^{(0)},\tilde{Q}_{{\bf n}}]$
determines the interaction between the fast and slow modes $Q_{{\bf n}%
}^{(0)} $, $\tilde{Q}_{{\bf n}}$ and has the form: 
\[
F_{int}[Q_{{\bf n}}^{(0)},\tilde{Q}_{{\bf n}}]=F_{kin}^{\prime }[Q_{{\bf n}%
}^{(0)},\tilde{Q}_{{\bf n}}]+F_{m}^{\prime }[Q_{{\bf n}}^{(0)},\tilde{Q}_{%
{\bf n}}] 
\]
\begin{eqnarray}
F_{kin}^{\prime }[Q_{{\bf n}}^{(0)},\tilde{Q}_{{\bf n}}] &=&\frac{\pi \nu }{2%
}{\rm Str}\int d{\bf r}d{\bf n}\biggl[Q_{{\bf n}}^{(0)}({\bf r})\bar{\tilde{T%
}}_{{\bf n}}({\bf r})v_{F}{\bf n\nabla _{r}}\tilde{T}_{{\bf n}}({\bf r}) 
\nonumber \\
&&+i\left( \frac{\omega +i\delta }{2}-\hat{a}({\bf r})\right) \tilde{T}_{%
{\bf n}}({\bf r})Q_{{\bf n}}^{(0)}({\bf r})\bar{\tilde{T}}_{{\bf n}}({\bf r})%
\biggr]  \nonumber
\end{eqnarray}
\begin{eqnarray}
F_{m}^{\prime }[Q_{{\bf n}}^{(0)}, &&\tilde{Q}_{{\bf n}}]=\left( \frac{\pi
\nu }{2}\omega _{c}\right) ^{2}\int d{\bf r}d{\bf n}d{\bf r^{\prime }}d{\bf %
n^{\prime }}W_{B}({\bf r}-{\bf r^{\prime }})  \nonumber \\
&&\times {\rm Str}\bigl[\hat{\tau}_{3}Q_{{\bf n}}^{(0)}({\bf r})\Phi _{{\bf n%
}}({\bf r})\bigr]{\rm Str}\bigl[\hat{\tau}_{3}Q_{{\bf n^{\prime }}}^{(0)}(%
{\bf r^{\prime }})\Phi _{{\bf n^{\prime }}}({\bf r^{\prime }})\bigr]+ 
\nonumber \\
&&+2\left( \frac{\pi \nu }{2}\omega _{c}\right) ^{2}\int d{\bf r}d{\bf n}d%
{\bf r^{\prime }}d{\bf n^{\prime }}W_{B}({\bf r}-{\bf r^{\prime }})
\label{eq26} \\
&&\times {\rm Str}\bigl[\hat{\tau}_{3}Q_{{\bf n}}^{(0)}({\bf r})\Phi _{{\bf n%
}}({\bf r})\bigr]{\rm Str}\bigl[\Lambda \hat{\tau}_{3}\bar{V}_{{\bf %
n^{\prime }}}({\bf r^{\prime }})i\partial _{\varphi ^{\prime }}V_{{\bf %
n^{\prime }}}({\bf r^{\prime }})\bigr]  \nonumber
\end{eqnarray}
\[
\Phi _{{\bf n}}({\bf r})=\bar{\tilde{T}}_{{\bf n}}({\bf r})i\partial
_{\varphi }\tilde{T}_{{\bf n}}({\bf r}) 
\]
Before the averaging over the fast fluctuations $Q_{{\bf n}}^{(0)}$ we make
the following essential remark.

The separation into the fast and slow modes, Eq.(\ref{eq23}), requires a
more accurate definition. The point is that the excitations in the model Eq.(%
\ref{eq22}) reveal a strong anisotropy in the phase space $({\bf r},{\bf n})$
due to the specific form of the free energy functional, Eq.(\ref{eq22}).
Since only the first order derivatives in ${\bf r}$ and ${\bf n}$ enter the
free energy, Eq.(\ref{eq22}), the dependence of the excitations on the
coordinates $({\bf r},{\bf n})$ will resemble a propagation along a
classical trajectory. Such an anisotropy demands a care and \ should be
performed in an invariant way. As in Ref. \cite{EK03}, the scale separation
can be performed introducing an additional term into the functional $F[Q_{%
{\bf n}}^{(0)}]$, Eq.(\ref{eq25}), 
\begin{equation}
F_{L}[Q_{{\bf n}}^{(0)}]=-\frac{\pi \nu }{2}\lambda _{L}{\rm Str}\int d{\bf r%
}d{\bf n}\Lambda Q_{{\bf n}}^{(0)}({\bf r})  \label{eq27}
\end{equation}
Then, we extend the region of the integration over $Q_{{\bf n}}^{(0)}({\bf r}%
)$ to all possible matrices with the constraints Eq.(\ref{eq18}). The
parameter $\lambda _{L}$ is just the Lyapunov exponent $\tau _{L}^{-1}$ and
the term $F_{L}[Q_{{\bf n}}^{(0)}]$, Eq.(\ref{eq27}), serves to suppress
fluctuations of the matrices $Q_{{\bf n}}^{(0)}$ outside the Lyapunov region.

As soon as the mode separation is properly defined one can carry out the
integration in Eq.(\ref{eq25}) and evaluate the effective energy $F_{eff}[%
\tilde{Q}_{{\bf n}}]$. We perform this computation using the cumulant
expansion in $F_{int}$, Eq.(\ref{eq25}) and approximation of the weak
magnetic field. In the same way as it was done in Ref. \cite{EK03} for the
model of the long-ranged disorder one can show that this is an expansion in
powers of the operator $l_{L}\nabla _{r}$ which is small outside the
Lyapunov region. Considering only the first order we find 
\begin{equation}
F_{eff}[\tilde{Q}_{{\bf n}}]=\langle F_{int}[Q_{{\bf n}}^{(0)},\tilde{Q}_{%
{\bf n}}]\rangle _{0}  \label{eq28}
\end{equation}
where the brackets $\langle \dots \rangle _{0}$ stand for integration over $%
Q_{{\bf n}}^{(0)}$. Due to the supersymmetry $\langle Q_{{\bf n}}^{(0)}({\bf %
r})\rangle _{0}=\Lambda $, which gives 
\begin{equation}
\langle F_{kin}^{\prime }[Q_{{\bf n}}^{(0)},\tilde{Q}_{{\bf n}}]\rangle
_{0}=F_{kin}[\tilde{Q}_{{\bf n}}]  \label{eq29}
\end{equation}
with the same functional $F_{kin}[\tilde{Q}_{{\bf n}}]$ as in Eq.(\ref{eq22}%
). The second term in the functional $F_{m}^{\prime }[Q_{{\bf n}}^{(0)},%
\tilde{Q}_{{\bf n}}]$ Eq.(\ref{eq26}) vanishes after the averaging due to
the symmetry as well. The contribution coming from the first term can be
divided into two parts: the first one comes from the reducible average and
coincides with the magnetic energy $F_{m}[\tilde{Q}_{{\bf n}}]$ of the
initial functional, Eq.(\ref{eq22}), whereas the other is given by the
irreducible average $\langle \langle Q_{{\bf n}}^{(0)}Q_{{\bf n^{\prime }}%
}^{(0)}\rangle \rangle _{0}=\langle Q_{{\bf n}}^{(0)}Q_{{\bf n^{\prime }}%
}^{(0)}\rangle _{0}-\langle Q_{{\bf n}}^{(0)}\rangle _{0}\langle Q_{{\bf %
n^{\prime }}}^{(0)}\rangle _{0}$ of the supermatrices $Q_{{\bf n}}^{(0)}$.

In order to find the contribution coming from the irreducible average we
consider the matrix 
\begin{equation}
\tilde{g}_{{\bf n}_{1}}({\bf r}_{1};\alpha )=\frac{\langle Q_{{\bf n}%
_{1}}^{(0)}({\bf r}_{1})\exp \left[ \frac{\pi \nu }{2}{\rm Str}\int d{\bf r}d%
{\bf n}\;\hat{a}_{{\bf n}}({\bf r})Q_{{\bf n}}^{(0)}({\bf r})\right] \rangle
_{0}}{\langle \exp \left[ \frac{\pi \nu }{2}{\rm Str}\int d{\bf r}d{\bf n}\;%
\hat{a}_{{\bf n}}({\bf r})Q_{{\bf n}}^{(0)}({\bf r})\right] \rangle _{0}}
\label{eq30}
\end{equation}
where the new source $\hat{a}_{{\bf n}}({\bf r})$ is 
\[
\hat{a}_{{\bf n}}({\bf r})=\alpha ({\bf r})\hat{\tau}_{3}\Phi _{{\bf n}}(%
{\bf r}),
\]
$\alpha ({\bf r})$ is some function. Due to the supersymmetry $\tilde{g}_{%
{\bf n}}({\bf r};\alpha =0)=\Lambda $. The first derivative in the function $%
\alpha ({\bf r})$ gives 
\begin{eqnarray}
\left. \frac{\delta \tilde{g}_{{\bf n}_{1}}({\bf r}_{1};\alpha )}{\delta
\alpha ({\bf r_{2}})}\right| _{\alpha ({\bf r})=0} &=&\frac{\pi \nu }{2}%
\langle \langle Q_{{\bf n}_{1}}^{(0)}({\bf r}_{1})  \label{eq31} \\
&&\times {\rm Str}\int d{\bf n^{\prime }}Q_{{\bf n^{\prime }}}^{(0)}({\bf r}%
_{2})\hat{\tau}_{3}\Phi _{{\bf n^{\prime }}}({\bf r}_{2})\rangle \rangle _{0}
\nonumber
\end{eqnarray}
On the other hand, the matrix $\tilde{g}_{{\bf n}}({\bf r};\alpha )$
satisfies the equation 
\[
v_{F}{\bf n\nabla _{r}}\tilde{g}_{{\bf n}}({\bf r};\alpha )+i\frac{\omega
+i\lambda _{L}}{2}\bigl[\Lambda ,\;\tilde{g}_{{\bf n}}({\bf r};\alpha )\bigr]%
=
\]
\begin{equation}
\alpha ({\bf r})\bigl[\hat{\tau}_{3}\Phi _{{\bf n}}({\bf r}),\;\tilde{g}_{%
{\bf n}}({\bf r};\alpha )\bigr]  \label{eq32}
\end{equation}
and condition $\tilde{g}_{{\bf n}}^{2}({\bf r};\alpha )=1$. Differentiating
in $\alpha ({\bf r})$ both sides of this condition and then putting $\alpha (%
{\bf r})=0$ we find that the matrix $\delta \tilde{g}_{{\bf n}}({\bf r}%
;\alpha )/\delta \alpha ({\bf r^{\prime }})|_{\alpha =0}$ in Eq.(\ref{eq31})
is off-diagonal. Eq(\ref{eq32}) can be considered for the off-diagonal part
of the matrix $\tilde{g}_{{\bf n}}({\bf r};\alpha )$ and rewritten in the
integral form 
\begin{equation}
\tilde{g}_{{\bf n}}^{\perp }({\bf r};\alpha )=\int d{\bf r^{\prime }}{\cal G}%
_{{\bf n}}({\bf r}-{\bf r^{\prime }})\alpha ({\bf r^{\prime }})\bigl[\hat{%
\tau}_{3}\Phi _{{\bf n}}({\bf r}),\;\tilde{g}_{{\bf n}}({\bf r};\alpha )%
\bigr]^{\perp }  \label{eq33}
\end{equation}
where the superscript $\bot $ stands for the part of the supermatrices
anticommuting with $\Lambda $. The kernel ${\cal G}_{{\bf n}}({\bf r}-{\bf %
r^{\prime }})$ is the solution of the equation 
\begin{equation}
\lbrack v_{F}{\bf n\nabla _{r}}+i(\omega +i\lambda _{L})\Lambda ]{\cal G}_{%
{\bf n}}({\bf r}-{\bf r^{\prime }})=\delta ({\bf r}-{\bf r^{\prime }})
\label{eq34}
\end{equation}
Differentiating in $\alpha ({\bf r})$ both sides of Eq. ({\ref{eq33}}) and
putting $\alpha ({\bf r})=0$ we obtain 
\[
\frac{\pi \nu }{2}\langle \langle Q_{{\bf n}_{1}}^{(0)}({\bf r}_{1}){\rm Str}%
\int d{\bf n^{\prime }}Q_{{\bf n^{\prime }}}^{(0)}({\bf r}_{2})\hat{\tau}%
_{3}\Phi _{{\bf n^{\prime }}}({\bf r}_{2})\rangle \rangle _{0}=
\]
\begin{equation}
{\cal G}_{{\bf n}_{1}}({\bf r}_{1}-{\bf r}_{2})\bigl[\hat{\tau}_{3}\Phi _{%
{\bf n}_{1}}({\bf r}_{2}),\;\Lambda \bigr]  \label{eq35}
\end{equation}
Substitution of Eq. (\ref{eq35}) into the Eq.(\ref{eq26}) gives 
\begin{eqnarray}
\langle F_{m}^{\prime }[\tilde{Q}_{{\bf n}},Q_{{\bf n}}^{(0)}]\rangle _{0}
&=&F_{m}[\tilde{Q}_{{\bf n}}]-\pi \nu \omega _{c}^{2}\int d{\bf r}d{\bf %
r^{\prime }}d{\bf n}W_{B}({\bf r}-{\bf r^{\prime }})  \nonumber \\
&&\times {\rm Str}\left[ \Phi _{{\bf n}}^{\bot }({\bf r}){\cal G}_{{\bf n}}(%
{\bf r}-{\bf r^{\prime }})\Lambda \Phi _{{\bf n}}^{\perp }({\bf r^{\prime }})%
\right]   \label{eq36}
\end{eqnarray}
Characteristic values of the difference ${\bf r}-{\bf r^{\prime }}$ in $%
{\cal G}_{{\bf n}}({\bf r}-{\bf r^{\prime }})$ are in the Lyapunov region,
whereas $\Phi _{{\bf n}}({\bf r})$ is a smooth function. This allows us to
make the replacement ${\bf r^{\prime }}\rightarrow {\bf r}$ in one of the $%
\Phi _{{\bf n}}$ in Eq.(\ref{eq36}). The integral over the difference ${\bf %
\rho }={\bf r}-{\bf r^{\prime }}$ is calculated as follows. First, we
rewrite this integral using integration in the momentum space instead of the
coordinate one 
\begin{equation}
\int {\cal G}_{{\bf n}}({\bf \rho })W_{B}({\bf \rho })d{\bf \rho }=\int 
\frac{d{\bf q}}{(2\pi )^{2}}W_{B}({\bf q})\frac{i}{v_{F}{\bf nq}-(\omega
+i\lambda _{L})\Lambda }  \label{eq37}
\end{equation}
The momentum ${\bf q}$ may be considered as the transfer momentum ${\bf q}=%
{\bf p^{\prime }}-{\bf p}$, where ${\bf p^{\prime }}=p_{F}{\bf n^{\prime }}$%
, ${\bf p}=p_{F}{\bf n}$ are momenta of a particle after and before the
scattering. Since for a weak scattering the characteristic length $b$ of the
distribution $W_{B}({\bf r}-{\bf r^{\prime }})$ is much smaller than the
Lyapunov length, $b\ll l_{L},$ Eq.(\ref{e3.1}), the fraction in Eq.(\ref
{eq37}) can be replaced by the $\delta $-function 
\[
\int \frac{d{\bf q}}{(2\pi )^{2}}W_{B}({\bf q})\frac{i}{v_{F}{\bf nq}%
-(\omega +i\lambda _{L})\Lambda }
\]
\begin{equation}
\approx -\pi \Lambda \int \frac{d{\bf q}}{(2\pi )^{2}}W_{B}({\bf q})\delta
(v_{F}{\bf nq})  \label{eq38}
\end{equation}
The $\delta $-function fixes the value of the final momentum ${\bf p^{\prime
}}$ on the Fermi surface: $\delta (v_{F}{\bf nq})=\delta \lbrack v_{F}{\bf n}%
({\bf p^{\prime }}-{\bf p})]=\delta \lbrack (\partial \varepsilon /\partial 
{\bf p})({\bf p^{\prime }}-{\bf p})]=\delta \lbrack \varepsilon ({\bf %
p^{\prime }})-\varepsilon ({\bf p})]$. Integrating over the energy $%
\varepsilon ^{\prime }\equiv \varepsilon ({\bf p^{\prime }})$ we find for
the integral, Eq.(\ref{eq37}), the following expression 
\begin{equation}
-\pi \nu \Lambda \int d{\bf n^{\prime }}W_{B}[p_{F}({\bf n}-{\bf n^{\prime }}%
)]  \label{eq39}
\end{equation}
Taking together Eqs.(\ref{eq36}), (\ref{eq37}), and (\ref{eq39}) we obtain
the free energy $F_{eff}[Q_{{\bf n}}]$ of the reduced $\sigma $-model 
\begin{equation}
F_{eff}[Q_{{\bf n}}]=F[Q_{{\bf n}}]+F^{\prime }[Q_{{\bf n}}],  \label{e390}
\end{equation}
\begin{eqnarray}
F[Q_{{\bf n}}] &=&\frac{\pi \nu }{2}\int d{\bf r}d{\bf n}{\rm Str}\biggl[%
\bar{T}_{{\bf n}}({\bf r})v_{F}{\bf n\nabla _{r}}T_{{\bf n}}({\bf r})+
\label{e391} \\
&&i\left( \frac{\omega +i\delta }{2}\Lambda -\hat{a}({\bf r})\right) Q_{{\bf %
n}}({\bf r})+\frac{1}{4\tau _{tr}}(\partial _{\varphi }Q_{{\bf n}})^{2}%
\biggr]  \nonumber
\end{eqnarray}
\begin{eqnarray}
F^{\prime }[Q_{{\bf n}}] &=&-\left( \frac{\pi \nu }{2}\omega _{c}\right)
^{2}\int d{\bf r}d{\bf n}d{\bf r^{\prime }}d{\bf n^{\prime }}W_{B}({\bf r}-%
{\bf r^{\prime }})  \label{e392} \\
&&\times {\rm Str}\left( \Lambda \hat{\tau}_{3}\bar{T}_{{\bf n}}({\bf r}%
)\partial _{\varphi }T_{{\bf n}}({\bf r})\right) {\rm Str}\left( \Lambda 
\hat{\tau}_{3}\bar{T}_{{\bf n^{\prime }}}({\bf r^{\prime }})\partial
_{\varphi ^{\prime }}T_{{\bf n^{\prime }}}({\bf r^{\prime }})\right)  
\nonumber
\end{eqnarray}
The collision term in the free energy functional is expressed through the
transport time $\tau _{tr}$ 
\begin{equation}
\left( 2\pi \nu \tau _{tr}\right) ^{-1}=\int d{\bf n^{\prime }}\omega
_{c}^{2}W_{B}[p_{F}({\bf n^{\prime }}-{\bf n})]  \label{eq41}
\end{equation}
and agrees with the results of the Refs.\cite{te1,te2,EK03} where the RMF
and long-range disorder models, respectively, were considered in the limit
of small scattering angles. The second term $F^{\prime }[Q_{{\bf n}}]$ in
Eq. (\ref{e390}) is small and can be neglected. This can be easily
understood using the fact that the Fourier transform of the function $W_{B}$
in Eq. (\ref{eq41}) contains momenta of the order of $p_{F}$, which
corresponds to short distances of the order of $\lambda _{F}$. In contrast,
the main contribution to the integral over the coordinates in Eq. (\ref{e392}%
) comes at weak RMF from larger distances of order $l_{L}$ where the
function $W_{B}$ is small. Therefore, everywhere below we will imply that
the reduced ballistic $\sigma $-model is described by the free energy
functional $F[Q_{{\bf n}}]$ from Eq. (\ref{e391}).

Thus, we have demonstrated that, although the ballistic $\sigma $-model for
the RMF is different from the one for the RP (the terms $F_{imp}[Q_{{\bf n}}]
$ and $F_{m}[Q_{{\bf n}}]$ in Eq. (\ref{eq22}) are different), the reduced $%
\sigma $-models describing the electron motion exceeding the Lyapunov length 
$l_{L}$ have the same form of Eq. (\ref{e391}). The similarity of the RMF
and RP models has been emphasized in Ref. \cite{gm1} and the final
conclusion of Ref. \cite{te2} was the same. However, the methods used in
these works were based on writing first the self-consistent Born
approximation for one particle Green functions (saddle point equation in the 
$\sigma $-model formulation) and on a subsequent expansion in slow modes,
which could not be justified at short distances. Now we see that the
equivalence of the RMF and RP models can hold at distances exceeding the
Lyapunov length. This naturally leads to the equivalence of the diffusive $%
\sigma $-models that can be written in the standard form 
\begin{equation}
F\left[ Q\right] =\frac{\pi \nu }{8}{\rm Str}\int \left[ D\left( \nabla
Q\right) ^{2}+2i\left( \omega +i\delta \right) \Lambda Q\right] d{\bf r}
\label{e500}
\end{equation}
where $D=v_{F}^{2}\tau _{tr}/2$. For the RMF problem the transport time $%
\tau _{tr}$ is given by Eq. (\ref{eq41}). 

Eq. (\ref{e500}) is valid unless the correlations of the magnetic field are
very long ranged. Only if 
\begin{equation}
\left\langle B_{q}B_{-q}\right\rangle \sim q^{-2}  \label{e501}
\end{equation}
an additional term can appear\cite{te2}. The symmetry of the diffusive $%
\sigma $-model, Eq. (\ref{e500}), corresponds to the unitary ensemble and
one comes to the standard conclusion about the localization.   

Of course, the coarse graining procedure leading to the ballistic $\sigma $%
-model, Eq. (\ref{e391}), is possible only if the ground state of the
initial $\sigma $-model, Eq. (\ref{eq22}), is achieved at $Q=\Lambda $.  One
can imagine such functions $W_{B}\left( {\bf r-r}^{\prime }\right) $ that
this ground state is no longer stable. However, this could be possible only
if the Fourier transform $W_{B}\left( {\bf q}\right) $ was negative for
certain ${\bf q}$, which is excluded in the case of real magnetic fields.
Therefore, beyond the Lyapunov region, the ballistic $\sigma $-model, Eq. (%
\ref{e391}), and, correspondingly, the diffusive $\sigma $-model, Eq. (\ref
{e500}),  seem to be unavoidable.

\section{Discussion}

\label{discussion}

In the present paper we considered the problem of the two-dimensional
electron gas in a random magnetic field (RMF) using the non-linear
supermatrix $\sigma $-model approach. We derived a ballistic $\sigma $-model
avoiding the standard scheme based on finding a saddle point in the integral
over supervectors and expanding in slow modes near this point. Such a scheme
explicitly relies on the assumption of a sufficiently short correlation
length of a random potential (see e.g. in Ref.\cite{E83}) and its validity
for a long range disorder is not clear. As the vector potential entering the
RMF model has a large correlation length even when the magnetic field is $%
\delta $-correlated in space, the procedure of singling out slow modes used
in the standard derivation is not well justified at least at not very large
distances. Besides, the saddle-point approximation is hardly allowed in this
case as well.

Instead of following the standard scheme we used the method based on writing
quasiclassical equations for Green functions and the exact representation of
their solutions in terms of integrals over supermatrices $Q_{{\bf n}}$ with
the constraint $Q_{{\bf n}}^{2}=1$. This method needs neither singling out
the ``fast'' and ``slow'' parts from the interaction nor the saddle-point
approximation. Conditions of the applicability of the method coincide with
those of the quasiclassical approximation. Therefore, the $\sigma $-model
obtained should be applicable over the distances down to the Fermi
wavelength, which makes it more general in comparison with the $\sigma $%
-models derived earlier on the basis of the standard scheme, Refs. \cite
{AMW94,aos,te1,te2}. The latter models are justified at distances exceeding
the single-particle mean free path $l$ as in the Ref.\cite{te1,te2} or the
transport length $l_{tr}$ as in Ref\cite{AMW94}.

We have demonstrated that similar to the problem of long range random
potential, there is a characteristic or Lyapunov length $l_{L}$
dividing the length scale into the Lyapunov and collision regions. The first
region corresponds to the small distances over which the particle motion is
strongly correlated. Correlations disappear over the larger lengths where
the particle interaction can be considered in terms of collisions. In the
Appendix we estimate the Lyapunov length for RMF problem restricting our
consideration by the limit of a weak field. The estimated length is
expressed through the transport length $l_{tr}$ and the correlation length $b
$ of the RMF by a formula similar to the one obtained previously in Ref.\cite
{AL96} in the model of a long ranged potential.

The reduced $\sigma $-model obtained after the integrating over the
fluctuations in the Lyapunov region coincides with the model of Ref.\cite
{te2} provided the latter is considered in the limit of a small angle weak
scattering. The reduced $\sigma $-model obtained in this way is equivalent
to the model found in the problem of a long range potential disorder Ref. 
\cite{EK03}. At the same time, it is relevant to emphasize that at short
distances inside the Lyapunov region the RMF and RP models correspond to
different $\sigma $-models. 

At distances, exceeding the transport length $l_{tr}=v_F\tau _{tr}$ one comes to the
standard diffusion $\sigma $-model, Eq. (\ref{e500}), unless the correlation
of the magnetic fields obeys Eq. (\ref{e501}). Calculations for the $\sigma $%
-model, Eq. (\ref{e500}), within the renormalization group scheme leads to
the standard conclusion about the localization. This conclusion is in
contradiction with the numerical results of Ref. \cite{N02} where the
existence of ``hidden degrees of freedom'' was proposed, which could lead to
the existence of extended states. We did not find any indication for such
degrees of freedom. Of course, our consideration was performed in the
quasiclassical limit, such that we did not take into account a possibility
of a quantization of the energy levels. However, it is not easy to
understand how taking into account distances shorter than the wave length $%
\lambda _{F}$ could lead to a destruction of the localization. 

\appendix

\section{%
\index{appendix}Lyapunov exponent in RMF problem}

Here we study the classical scattering of two particles in a random magnetic
field (RMF). The presence of the RMF leads to an effective interaction
between the particles. The radius of this interaction is equal to the
correlation length of the field. The scattering process lasts a finite time
after which the particles diverge over the distance exceeding the
correlation length and begin to move without any interaction. The aim of the
calculation presented below is to estimate this time. It is clear that for
larger times the particle scattering may be considered in terms of
collisions. We restrict our calculation by the case of a weak magnetic field.

Let us consider two particles on a plane with the coordinates ${\bf r}_{1}$, 
${\bf r}_{2}$ and momenta ${\bf p}_{1}$, ${\bf p}_{2}$ moving in a
perpendicular magnetic field. The equations of the motion for each particle
are 
\begin{equation}
\dot{{\bf r}_{i}}=\frac{{\bf p}_{i}}{m},\text{ \ \ }\dot{{\bf p}_{i}}=\frac{%
eB({\bf r}_{i})}{mc}[{\bf p}_{i}\times \hat{e}_{z}]  \label{a1}
\end{equation}
where $\hat{e}_{z}$ is the unit vector perpendicular to the plane of the
motion. Let ${\bf \rho }={\bf r}_{1}-{\bf r}_{2}$ and ${\bf p}={\bf p}_{1}-%
{\bf p}_{2}$ be coordinate and momentum of the relative motion. We assume
that the particles start their motion close to each other and have parallel
momenta ${\bf p}_{1}={\bf p}_{2}$ so that ${\bf p}=0$ and ${\bf \rho }={\bf %
\rho }_{0}$ in the beginning; ${\bf \rho }_{0}$ is assumed to be
perpendicular to the direction ${\bf n}$ of the motion of the center of
mass. Since the energy does not change in the magnetic field, the absolute
value of the momenta ${\bf p}_{1},{\bf p}_{2}$ will remain constant and
equal to each other $|{\bf p}_{1}|=|{\bf p}_{2}|$. Therefore, the direction
of the relative motion will always be perpendicular to the direction of the
motion of the mass center ${\bf n}$: $({\bf pn})=0$. This allows us to write 
${\bf \rho }=\rho \lbrack {\bf n}\times \hat{e}_{z}]$. Using Eq.(\ref{a1})
we find 
\begin{equation}
\dot{\rho}=\frac{p}{m},\text{ \ \ }\dot{p}=e\frac{v_{F}}{c}(B_{1}-B_{2})
\label{a2}
\end{equation}
where $B_{i}\equiv B({\bf r}_{i})$, and $p=|{\bf p}|$ is the absolute value
of the momentum of the relative motion. At the beginning of the motion $\rho 
$ is rather small and the difference $B_{1}-B_{2}$ can be approximately
written as $B_{1}-B_{2}\approx (\partial B/\partial R_{\perp })\rho $, where 
$R_{\perp }$ is the coordinate of the mass center in the direction
perpendicular to ${\bf n}$. Eq.(\ref{a2}) considered in this approximation
reduces to a linear system of first order differential equations. Hence, the
distance $\rho $ will grow exponentially as a function of time. The mean
rate of the divergency or the Lyapunov exponent determines the scattering
time involved.

To study statistics of the relative motion we introduce a distribution
function $W(t,\rho ,p)$. By definition, it is the probability for the
relative distance and momentum to be $\rho $ and $p$ at the time $t$,
respectively, provided they have been initially $\rho _{0}$, $p=0$. Let $%
W(t_{0},\rho ,p)$ be the distribution at the time $t_{0}$. Then, it can be
written at the time $t_{0}+\Delta t$ as 
\begin{eqnarray}
W(t_{0}+\Delta t,\rho ,p) &=&\int P(t_{0}+\Delta t,\rho ,p;t_{0},\rho
^{\prime },p^{\prime })  \nonumber \\
&&\times W(t_{0},\rho ^{\prime },p^{\prime })d\rho ^{\prime }dp^{\prime }
\label{a3}
\end{eqnarray}
where $P(t,\rho ,p;t^{\prime },\rho ^{\prime },p^{\prime })$ is the
transition probability. This probability is determined by the equation of
motion, Eq. (\ref{a2}), and is introduced as 
\begin{eqnarray}
&&P(t,\rho ,p;t^{\prime },\rho ^{\prime },p^{\prime })=\delta \left( \rho
-\rho ^{\prime }-\int_{t^{\prime }}^{t}\frac{p(\tau )}{m}d\tau \right)  
\nonumber \\
&&\times \delta \left( p-p^{\prime }-e\frac{v_{F}}{c}\int_{t^{\prime
}}^{t}[B_{1}(\tau )-B_{2}(\tau )]d\tau \right)   \label{a4}
\end{eqnarray}
where $B_{i}(\tau )=B[{\bf r}_{i}(\tau )]$, ${\bf r}_{i}(\tau )={\bf R}(\tau
)\pm {\bf \rho }(\tau )/2$ and $\rho (\tau )$, $p(\tau )$ are the solution
of the classical motion equation (\ref{a2}). Substitution of Eq.(\ref{a4})
into Eq.(\ref{a3}) gives a relation between the distributions $W$ at times $%
t_{0}$ and $t_{0}+\Delta t$. Assuming that $\Delta t$ is smaller than the
inverse Lyapunov exponent we expand this relation in $\Delta t$ and then
average over the magnetic field $B({\bf r})$. Since the magnetic field is
assumed to be weak, we neglect the influence of the field on the trajectory
of the mass center and obtain 
\begin{equation}
\frac{\partial W}{\partial t}+\frac{p}{m}\frac{\partial W}{\partial \rho }-%
\frac{2}{\tau _{tr}}\varepsilon ({\rho })\;p_{F}^{2}\frac{\partial ^{2}W}{%
\partial p^{2}}=0  \label{a6}
\end{equation}
where $\tau _{tr}$ is the transport time, Eq.(\ref{eq41}), that can also be
written as 
\begin{equation}
\frac{1}{\tau _{tr}}=\omega _{c}^{2}\int_{-\infty }^{+\infty }W_{B}(v_{F}%
{\bf n}\tau )d\tau   \label{a7}
\end{equation}
The function $\varepsilon (\rho )$ is by definition 
\begin{equation}
\varepsilon (\rho )=1-\frac{\int_{-\infty }^{+\infty }W_{B}(v_{F}{\bf n}\tau
+\rho \lbrack {\bf n}\times \hat{e}_{z}])d\tau }{\int_{-\infty }^{+\infty
}W_{B}(v_{F}{\bf n}\tau )d\tau }  \label{a8}
\end{equation}
The distance between the particles in the Lyapunov region is smaller than
the correlation length of the magnetic field $b$. Hence, one may expand the
function $\varepsilon (\rho )$ in $\rho $, which gives $\varepsilon (\rho
)\approx \rho ^{2}/2b^{2}$. This relation is to be considered as a
definition of the length $b$. Substituting this expansion into Eq.(\ref{a6})
we come to the same equation as the one derived in Ref.\cite{AL96} where
electron scattering in a long-ranged potential disorder was considered. 
\begin{equation}
\left[ \frac{\partial }{\partial t}-v_{F}\phi \frac{\partial }{\partial \rho 
}-\frac{\rho ^{2}}{\tau _{tr}a^{2}}\frac{\partial ^{2}}{\partial \phi ^{2}}%
\right] W=0  \label{a80}
\end{equation}

Using the result of that paper we find that the function $W(t,\rho )$
determining the distribution of the distance $\rho $ (the momentum of the
relative motion ${\bf p}$ is implied to be averaged in this function)
satisfies the equation 
\begin{equation}
\left[ \tau _{L}\frac{\partial }{\partial t}-\beta \frac{\partial }{\partial
z}\right] W=0  \label{a9}
\end{equation}
where $\beta $ is a numerical coefficient equal to $\beta \approx 0.365$ and 
$z=\ln (b/\rho )$. It follows from Eq.(\ref{a9}) that the coefficient $\tau
_{L}$ is in fact a characteristic time of the divergency of the trajectories
of the particles calculated from the classical motion equation Eq.(\ref{a2}%
). According to Ref.\cite{AL96} this time is equal to 
\begin{equation}
\tau _{L}=\tau _{tr}\left( \frac{b}{l_{tr}}\right) ^{2/3}  \label{a10}
\end{equation}
and this is at the same time the inverse Lyapunov exponent. As mentioned
above, the quantity $\tau _{L}$ has the meaning of a characteristic time
that two scattered particles spend moving together until the distance
between them starts exceeding the correlation length $b$.

\widetext

\end{document}